\documentclass[letter,superscriptaddress,twocolumn, prb,showkeys]{revtex4-1}
	\usepackage{amsmath, amssymb} 
	\usepackage{makeidx}
	\usepackage{amsfonts}
	\usepackage{dsfont}
	\usepackage[ansinew]{inputenc}
	\usepackage[usenames,dvipsnames]{pstricks}
	\usepackage{epsfig}
	\usepackage{pst-grad} 
	\usepackage{pst-plot} 
	\usepackage[colorlinks,hyperindex]{hyperref}
	\usepackage{graphicx}
	\usepackage{caption} 
	\usepackage{subcaption}
	\hypersetup
	{
		colorlinks,%
		citecolor=black,%
		linkcolor=black,%
		urlcolor=black,%
	}

	\newcommand{\CN}{{\cal N}}
    
    \newcommand{\CO}{{\cal O}}
    \newcommand{\CP}{{\cal P}}
    \newcommand{\CR}{{\cal R}}
	\newcommand{\vev}[1]{{\left< {#1} \right>}}
	
\makeindex

\begin{document}

\title{The Exact Bremsstrahlung Function in $\CN=2$ Superconformal Field Theories}

\author{Bartomeu Fiol}
\email{bfiol@ub.edu}
\affiliation{Departament de Física Fonamental, Institut de Ci\`encies del Cosmos (ICCUB), \\
Universitat de Barcelona, Mart\'i i Franqu\`es 1, 08028 Barcelona, Spain}
\author{Efrat Gerchkovitz}
\email{efrat.gerchkovitz@weizmann.ac.il}
\author{Zohar Komargodski}
\email{zohar.komargodski@weizmann.ac.il}
\affiliation{Department of Particle Physics and Astrophysics \\ Weizmann Institute of Science, Rehovot 76100, Israel}


\begin{abstract}
 We propose an exact formula for the energy radiated by an accelerating quark in $\CN=2$ superconformal theories in four dimensions. This formula reproduces the known Bremsstrahlung function for $\CN=4$ theories and provides a prediction for all the perturbative and instanton corrections in $\CN=2$ theories. We perform a perturbative check of our proposal up to three loops.  
\end{abstract}

\keywords{}

\maketitle


\section{Introduction and Review}

Many interesting questions in Quantum Field Theory revolve around the behavior of external probes coupled to the theory. In particular, if a heavy particle moves with some proper acceleration $a$ in the vacuum of a gauge theory, it radiates energy proportional to the proper acceleration squared, 
\begin{equation}\label{Bdef}
E=2\pi B\int dt \;a^2~.
\end{equation}
The well known result (Larmor's formula) for a particle of charge $e$ in Maxwell's theory is 
\begin{equation}\label{Larmor}B={e^2\over 12\pi^2}~.\end{equation}

A convenient general way to describe a charged heavy probe is by a Wilson operator. It is labeled by the representation $\CR$ of the gauge group and worldline C. To discuss energy loss, we start from the probe being at rest and then it receives a sudden kick, continuing thereafter at a constant speed. The worldline thus has a cusp, and the vacuum expectation value of the Wilson operator develops a logarithmic divergence that depends on the boost parameter $\varphi$ 
\begin{equation}
\vev{W_\varphi}\sim e^{-\Gamma_{\text{cusp}}(\varphi)\log{{\Lambda_{UV}}\over{\Lambda_{IR}}}}\;,
 \end{equation} where $\Lambda_{UV}$ and $\Lambda_{IR}$ represent UV and IR cut-off scales.\cite{Polyakov:1980ca} 
The quantity $\Gamma_{\text{cusp}}(\varphi)$ is the cusp anomalous dimension and it plays an important role in a number of questions, like the IR divergences in the scattering of massive particles. It has been computed to three-loops in QCD \cite{Grozin:2014hna}  and in ${\cal N}$=4 SYM \cite{Correa:2012nk}, and to four-loops in planar $\CN=4$.\cite{Henn:2013wfa}

While obtaining the full expression for $\Gamma_{\text{cusp}}(\varphi)$ in any interacting gauge theory appears to be a daunting task, various limits of this function are more accessible, and already encode interesting physics. In what follows we will limit the discussion to conformal field theories, although some of the results are more general. In the limit of very large boosts, $\Gamma_{\text{cusp}}(\varphi)$ is linear in the boost parameter \cite{Korchemsky:1987wg, Korchemsky:1991zp}, 
\begin{equation}
\Gamma_{\text{cusp}}(\varphi)\sim \Gamma_{\text{cusp}}^\infty \varphi
\end{equation}
and characterizes the IR divergences of massless particles. On the other hand, in the limit of very small boosts we have,
\begin{equation}
\Gamma_{\text{cusp}}(\varphi)=B\varphi^2+{\cal O}(\varphi^4)~.
\label{brem}
\end{equation}
The coefficient $B$ was dubbed the Bremsstrahlung function in \cite{Correa:2012at}, where it was argued that for conformal field theories it determines the energy radiated by an accelerating quark, as in (\ref{Bdef}). It also captures the momentum diffusion coefficient of the accelerated probe. \cite{Fiol:2013iaa}

Let us now discuss the Wilson line corresponding to the trajectory of a probe moving at constant proper acceleration. We can measure the energy density by studying the two-point function of the stress-energy tensor and this Wilson line. In conformal field theories, this is related by a conformal transformation to the two-point function of the stress-energy tensor and a straight Wilson line,
\begin{equation}\label{twopoint}
\vev{T_{\mu \nu}(x)}_W \equiv \frac{\vev{WT_{\mu \nu}(x)}}{\vev{W}}~.
\end{equation}
Its $x$ dependence is determined by conformal invariance, up to a single coefficient $h_W$, \cite{Kapustin:2005py}
\begin{equation}
\vev{T_{00}(x)}_W=\frac{h_W}{r^4}~,
\label{defofh}
\end{equation}
where $r$ is the distance from the line. 
There is no simple general relation between $B$ and $h_W$.~\cite{Lewkowycz:2013laa}

The main subject of this paper is the computation of $B$ in $\mathcal{N}=2$ Superconformal Field Theories (SCFTs). We first review the case of maximally supersymmetric $\mathcal{N}=4$ SCFT.

\subsection{Review of $\mathcal{N}=4$}
The massive probe that we are studying is described by the Wilson loop in a representation ${\cal R}$ of the gauge group  
\begin{equation}
W_{\cal R}={1\over{\hbox{dim} {\cal R}}}\hbox{tr}_{\cal R}  \CP \, \hbox{exp} \;i\int  \left(A_\mu dx^\mu+i\Phi_i\theta^i ds \right)\;.\label{WilsonN=4}
\end{equation}
Here, $A_\mu$ and $\Phi^i$, $i=1,..,6$ are the gauge fields and scalars of the $\CN=4$ vector multiplet, 
$\theta^i$ is some unit vector in $\mathbb{R}^6$ and $\CP$ is the path ordering operator.  When the contour is a straight line and $\theta^i$ is constant, $W_{\cal R}$ is $1/2$ BPS. Another 1/2 BPS configuration is given by a circular Wilson loop with constant $\theta^i$. The two configurations are formally related by a conformal transformation. 

For the straight line we have $\vev{W}=1$.\footnote{From now on we suppress the dependence on the representation $\CR$ of the gauge group.}
The transformation that relates the straight and circular Wilson loops turns out to be anomalous\cite{Drukker:2000rr} and as a result the expectation value of the circular Wilson loop is a non-trivial function of the coupling constant and the number of colors $N$. It was conjectured by Erickson-Semenoff-Zarembo~\cite{Erickson:2000af} and Drukker-Gross~\cite{Drukker:2000rr} and later proved by Pestun~\cite{Pestun:2007rz} using supersymmetric localization on $\mathbb{S}^4$ that the expectation value of the circular Wilson loop is given by a Gaussian matrix integral over the Lie algebra

\begin{equation}\label{intro}
\vev{W}=\frac{ \int  da\,  \hbox{Tr}\, e^{-2\pi  a}\, e^{-\frac{8\pi^2N}{\lambda}\hbox{Tr}( a^2)}}
{\int  d a \,  e^{-\frac{8\pi^2N}{\lambda}\,\hbox{Tr}(a^2) }}\;,
\end{equation}
where $\lambda=g^2N$ is the 't Hooft coupling, with $g$ the usual Yang-Mills coupling.

In~\cite{Correa:2012at} it was argued that the Bremsstrahlung function for $\CN=4$ $U(N)$ SYM can be derived from the vacuum expectation value of the 1/2 BPS circular Wilson loop by a derivative with respect to the 't Hooft coupling:
\begin{equation}
B=\frac{1}{2\pi^2}\lambda\partial_\lambda \hbox{ln } \vev {W}~.
\label{BinN=4}
\end{equation} 
In the 't Hooft limit and at large $\lambda$, this agrees with the replacement rule $e^2/3\leftrightarrow \sqrt \lambda$ found via the AdS/CFT correspondence.~\cite {Kruczenski:2002fb, Mikhailov:2003er} 

On the other hand, the coefficient $h_W(\lambda)$ for the two-point function of the stress-energy tensor and the $1/2$~BPS Wilson line was computed in~\cite{Fiol:2012sg}, obtaining a result equal to $B$, up to a numerical coefficient. This relation was further clarified in \cite{Lewkowycz:2013laa}, who argued for $\CN=4$ theories that 
\begin{equation}
B=3h_W\;.\label{B=3h}
\end{equation}
The argument relies on the existence of a dimension-two scalar operator in the supermultiplet of the energy momentum tensor. 

\subsection{Some Basics of $\mathcal{N}=2$}
Let us now consider $\CN=2$ SCFTs in four dimensions. We can define the following Wilson loop 
\begin{equation}\label{BPSN2}
W_{\cal R}={1\over{\hbox{dim} {\cal R}}}\hbox{tr}_{\cal R}  \CP \, \hbox{exp} \;i\oint  \left(A_\mu dx^\mu+i\Phi ds \right)~,
\end{equation}
with $\Phi$ one of the scalar fields in the ${\cal N}=2$ vector multiplet.
As before, if the contour is straight or circular, the Wilson loop will be 1/2 BPS. If we introduce a cusp then we can infer the Bremsstrahlung coefficient according to~\eqref{brem}.

The expectation value of the circular Wilson loop in ${\CN=2}$ SCFTs can be obtained via localization~\cite{Pestun:2007rz} on the four sphere, $\mathbb{S}^4$. Actually, for our purposes below, it is also useful to review what happens when the Wilson loop is placed on a squashed four-sphere. Consider the  ellipsoid:
\begin{equation}\label{back}
\frac{x_0^2}{r^2}+\frac{x_1^2+x_2^2}{\ell^2}+\frac{x_3^2+x_4^2}{\tilde \ell^2}=1\;.
\end{equation}
In SCFTs, the expectation value of the Wilson loop is now a function of the dimensionless squashing parameter 
\begin{equation}
b\equiv\left(\frac{\ell}{\tilde \ell}\right)^{1/2}\;.
\end{equation}
If we take $b=1$ we are back to the round $\mathbb{S}^4$, which is conformally equivalent to flat space (and we can thus extract the usual expectation value of the circular loop).

In~\cite{Hama:2012bg} Hama and Hosomichi computed the expectation value of the circular Wilson loop~\eqref{BPSN2} placed on the ellipsoid~\eqref{back}, see also \cite{Alday:2009fs, Fucito:2015ofa}. There are actually two supersymmetric Wilson loops on the ellipsoid. They transform into each other under ${\ell \leftrightarrow\tilde \ell}$, and approach the 1/2 BPS Wilson loop considered by Pestun in the round $\mathbb{S}^4$ limit $\ell=\tilde \ell=r$.
The vacuum expectation value of one of them is
\begin{equation}
\begin{aligned}
&\vev{W_b}=\\&\frac{ \int  da\,  \hbox{Tr}\, e^{-2\pi b a}\, e^{-\frac{8\pi^2}{g^2}\hbox{Tr}( a^2)}Z_{\text{1-loop}} ( a,b) \, |Z_{\text{inst}} ( a,b)|^2}
{\int  d a \,  e^{-\frac{8\pi^2}{g^2}\,\hbox{Tr}( a^2)}Z_{\text{1-loop}} ( a,b) \, |Z_{\text{inst}} (a,b)|^2}\;,
\end{aligned}\label{localization_eq}
\end{equation} 
while the second Wilson loop is obtained by replacing
 $\hbox{Tr}\, e^{-2\pi b
 	a}$ by $\hbox{Tr}\, e^{-2\pi b^{-1}a}$.
The integration in (\ref{localization_eq}) is, as before, over the Lie algebra. $Z_{\text{inst}}$ is Nekrasov's instanton partition function \cite{Nekrasov:2002qd} on the Omega background, with the equivariant parameters identified as
\begin{equation}
\ell=\epsilon_1^{-1}\hspace{1cm}\tilde \ell=\epsilon_2^{-1}\;,
\end{equation}
thus, $b\equiv\left(\frac{\epsilon_2}{\epsilon_1}\right)^{1/2}$. The expression for the 1-loop determinant, $Z_{\text{1-loop}}$, can be found in \cite{Hama:2012bg} (see also \cite{Pestun:2014mja}).

Consider now the normalized two-point function of the stress-energy tensor with a straight Wilson line~\eqref{twopoint} in an $\mathcal{N}=2$ SCFT. We get some function of the marginal coupling constants~\eqref{defofh}, $h_W(g^i)$. The marginal coupling constants couple to some chiral primary operators which are unrelated to the stress-energy multiplet. The stress-energy tensor belongs to a short representation of the $\CN=2$ superconformal group \cite{Dolan:2002zh} that always contains a scalar of dimension two, $O_2$ \cite{Dolan:2002zh} (but no scalars of dimension four). Because the Wilson loop is BPS, there is a simple relation between $\vev{WT_{\mu \nu}(x)}$ and $\vev{WO_2(x)}$. Namely, if we define $\vev{O_2(x)}_W=\frac{C}{r^2}$, then ${h_W(g^i)=\frac{8}{3}C(g^i)}$. This relation is derived in appendix A and it will be useful below.

\section{Two Conjectures}

Because of the relation just outlined between $\vev{WT_{\mu \nu}(x)}$ and $\vev{WO_2(x)}$, one can imagine, as in~\cite{Lewkowycz:2013laa}, improving the energy-momentum tensor in such a way that the leading singularity near the Wilson line is removed. This is possible in all $\mathcal{N}=2$ theories because an operator of dimension two always exists, and its correlation function with the Wilson line has precisely the same coupling constant dependence as the two-point function of the energy-momentum tensor and the Wilson line.

Therefore, we suggest that the Bremsstrahlung coefficient in $\mathcal{N}=2$ theories can be inferred from $h_W$ as in~\eqref{B=3h}
\begin{equation}\label{B2h_Wagain}
B=3h_W~.
\end{equation}

In general, $\mathcal{N}=2$ theories contain many exactly marginal operators and one should not expect a formula analogous to~\eqref{BinN=4} because these exactly marginal operators are unrelated to insertions of the energy-momentum tensor. Instead, we conjecture that the coefficient $h_W$ and therefore the Bremsstrahlung function for $\CN=2$ SCFTs is given by
 \begin{align}
 B =3h_W= \frac{1}{4\pi ^2} \partial_b \ln \, \vev{W_b} |_{b=1}\;.
 \label{theBguess}
 \end{align}

The proposal $h_W= \frac{1}{12\pi ^2} \partial_b \ln \, \vev{W_b} |_{b=1}$ is motivated by the fact that an infinitesimal equivariant deformation of $\mathbb{S}^4$ corresponds to an insertion of an integrated energy-momentum supermultiplet.\footnote{Analogous ideas have appeared, for example, in~\cite{Closset:2012ru,Lewkowycz:2013laa}.}

In the absence of the Wilson loop the background~\eqref{back} is invariant under $\epsilon_1 \leftrightarrow \epsilon_2$ and therefore the classical, one-loop and, instanton contributions start deviating from their $\mathbb{S}^4$ expressions only at second order in $b-1$. The Wilson loop insertion $\hbox{Tr}\, e^{-2\pi b  a}$ in equation~\eqref{localization_eq} is the only factor in the integrand that contains a term linear in $b-1$.  Therefore,
 $\vev{W_b}$ in (\ref{theBguess}) can be computed using just the 1-loop determinant and instanton factors of the round $\mathbb{S}^4$ matrix model.

It is worth pointing out that for planar ${\cal N}=2$ superconformal gauge theories, there is an interesting proposal \cite{Pomoni:2013poa, Mitev:2014yba} to obtain $\Gamma_{\text{cusp}}^\infty$ from the corresponding quantity in planar ${\cal N}=4$ SYM, by applying a substitution rule for the coupling. It would be interesting to see if that procedure also generalises for the coefficient~$B$.

\section{Tests of the Conjectures}

In the rest of the paper we provide some checks of the conjecture (\ref{theBguess}). For $\CN=4$ theories, we show that (\ref{theBguess}) is equivalent to (\ref{BinN=4}). For $\CN=2$ SCFTs, (\ref{theBguess}) predicts a deviation from the $\CN=4$ result starting at the $g^6$ order in perturbation theory. Indeed, we find that conformal invariance ensures that the one- and two-loop contributions to $h_W$ and $\Gamma_{\text{cusp}}$  are independent of the matter content.   For $SU(2)$ with four fundamental hypermultiplets, we compute the $g^6$ correction to $\Gamma_{\text{cusp}}$ and we find agreement with (\ref{theBguess}).  In addition, we show (for $SU(N)$ with $2N$ fundamental hypermultiplets) that the right hand side of (\ref{theBguess}) is positive, as required by the interpretation of $B$ as the energy radiated by a quark. 

\subsection{${\cal N}=4$}
For ${\cal N}=4$ U(N) SYM, it was proven in \cite{Correa:2012at} that
\begin{equation}
B=\frac{1}{2\pi^2}\lambda\partial_\lambda \hbox{ln } \vev {W}\;.
\end{equation}
Let us check that this is in agreement with our conjecture~(\ref{theBguess}).
The localization formula gives:
\begin{equation}
\vev{W_b}=\frac{ \int  da\,  \hbox{Tr}\, e^{-2\pi b a}\, e^{-\frac{8\pi^2N}{\lambda}\hbox{Tr}( a^2)}}
{\int  d a \,  e^{-\frac{8\pi^2N}{\lambda}\,\hbox{Tr}(a^2) }}+\CO\left((b-1)^2\right)\;.
\end{equation}
The rescaling of the integration variable $a = \sqrt{\lambda}\tilde a$ makes it manifest that $\vev{W_b}$ is a function of a single variable $b\sqrt\lambda$:
\begin{equation}
\vev{W_b}=\frac{ \int  d\tilde a\,  \hbox{Tr}\, e^{-2\pi b \sqrt{\lambda}\tilde a}\, e^{-{8\pi^2}N\hbox{Tr}( \tilde a^2)}}
{\int  d \tilde a \,  e^{-{8\pi^2}N\hbox{Tr}( \tilde a^2)}}+\CO\left((b-1)^2\right)\;.
\end{equation}
Thus, the conjectured formula~\eqref{theBguess} follows.

\subsection{Free $\mathcal{N}=2$ $U(1)$ theory}
The simplest  ${\cal N}=2$ SCFT is the free Abelian ${\cal N}=2$ gauge theory (with no matter). From the field theory side, the value of $h_W$ is the same as for the free Abelian ${\cal N}=4$ SYM. In the matrix model computation, the instanton contribution is now different from the identity \cite{Pestun:2007rz}, but since it is moduli independent, it pulls out of the integrals, and cancels out. Therefore, our conjecture (\ref{theBguess})  applies.

\subsection{$B$ and $h_W$ to two-loop order}

We now study non-trivial, perturbative, ${\cal N}=2$ SCFTs. The vanishing of the $\beta$-function implies that if we have $n_{\CR}$ hypermultiplets in the representation $\CR$ of the gauge group then \begin{equation}
C(Adj)=\sum_{\CR} n_{\CR} C(\CR)~.
\label{betavanish}
\end{equation}
As already noted in~ \cite{Pestun:2007rz}, this has the following implication for the one-loop determinant that appears in the matrix model: 
\begin{equation}
Z_{\text{1-loop}}(a)=1+{\cal O}(a^4)~,
\end{equation}
{i.e.} there is no ${\cal O}(a^2)$ term.  As a consequence, for ${\cal N}=2$ SCFTs,  the perturbative expansion of ${\vev W}$ starts depending on the matter content of the theory at order $g^6$. If the conjectured formula (\ref{theBguess}) is correct, the same thus must be true for the coefficients $B$ and $h_W$.

We begin by considering $h_W$, which is given by $\vev{O_2(x)W}$, where $O_2(x)$ is the superconformal primary in the supermultiplet of $T_{\mu\nu}(x)$. 
The strategy, as in \cite{Andree:2010na}, is to focus on the diagrams where the hypermultiplets enter,  and argue that by virtue of~(\ref{betavanish}) the result does not depend on the matter content. At order $g^2$ the hypermultiplets do not enter the computation, so the claim readily follows. 
At order $g^4$,  hypermultiplets fields appear only in the diagrams shown in figure~1. For each one of these diagrams, the dependence on $\{n_{\CR}\}$ is through the combination $\sum_{\CR} n_\CR C(\CR)$. Due to (\ref{betavanish}) this is independent of the matter content. 

Carrying out a similar study of the diagrams contributing to the cusped Wilson line
 $\Gamma_{\text{cusp}}$ up to order $g^4$ (see also\cite{Makeenko:2006ds}), we find the following:
 At order $g^2$ , the diagrams that contribute do not involve the hypermultiplet fields (figure 2a). At order $g^4$, hypermultiplet fields enter in the one-loop correction of the vector multiplet scalar and vector fields propagators (figure 2b). As before, the dependence of these diagrams on the number of hypermultiplets is of the form $\sum_{\CR} n_\CR C(\CR)$, which is equal to $C(Adj)$ for conformal theories. Thus, we can conclude that, for $\CN=2$ SCFTs, $B$ does not depend on the matter content up to the order $g^4$. 

Since our proposal~(\ref{theBguess}) gives the correct result for ${\CN=4}$, it follows that the conjectured formula~(\ref{theBguess}) is correct up to the order  $g^4$ in all $\mathcal{N}=2$ SCFTs.

\begin{figure}\label{fig:O2}
	\begin{minipage}{.3\linewidth}
		\includegraphics[width=\linewidth]{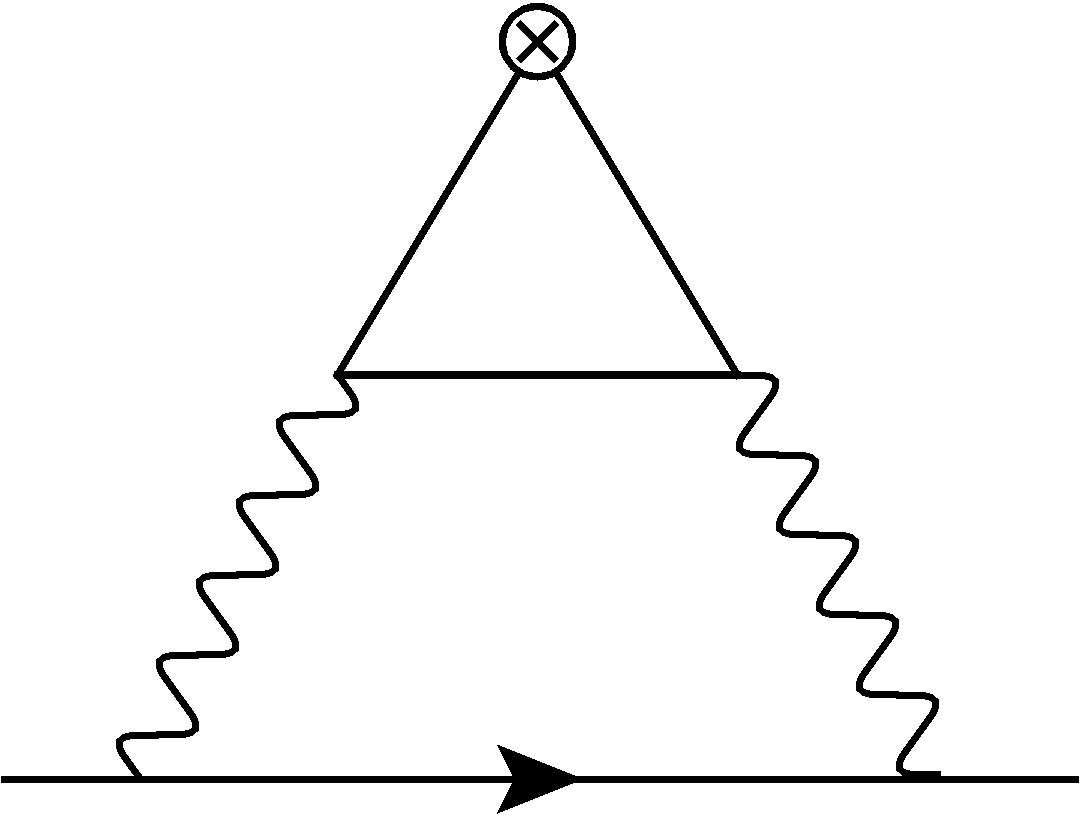}
		\label{img1}
	\end{minipage}
	\hspace{.05\linewidth}
	\begin{minipage}{.3\linewidth}
		\includegraphics[width=\linewidth]{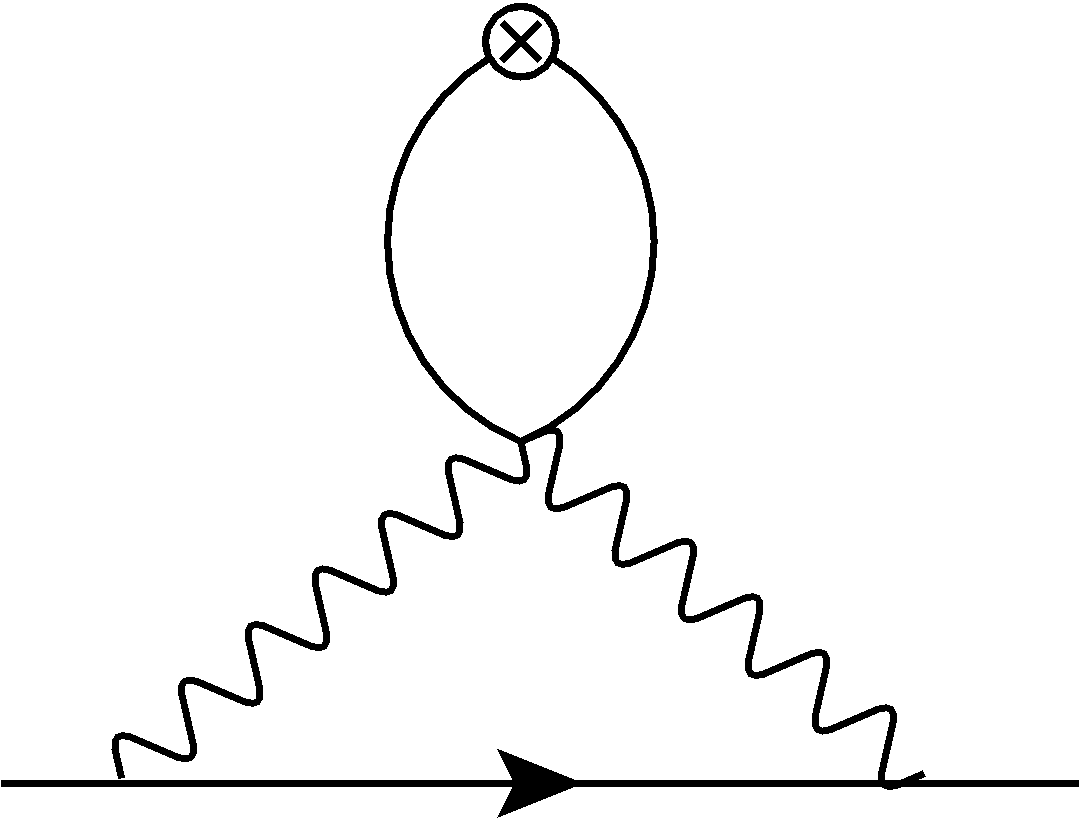}
		\label{img2}
	\end{minipage}
	\\
	\begin{minipage}{.3\linewidth}
		\includegraphics[width=\linewidth]{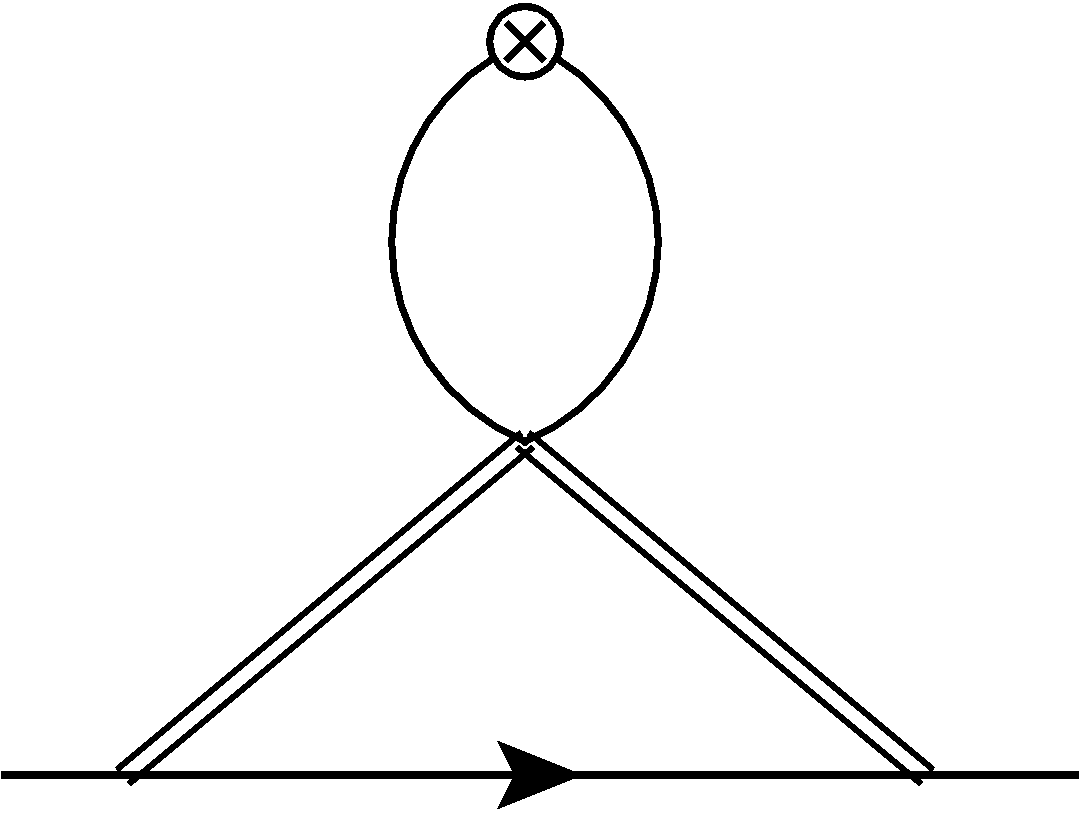}
		\label{img1}
	\end{minipage}
	\hspace{.05\linewidth}
	\begin{minipage}{.3\linewidth}
		\includegraphics[width=\linewidth]{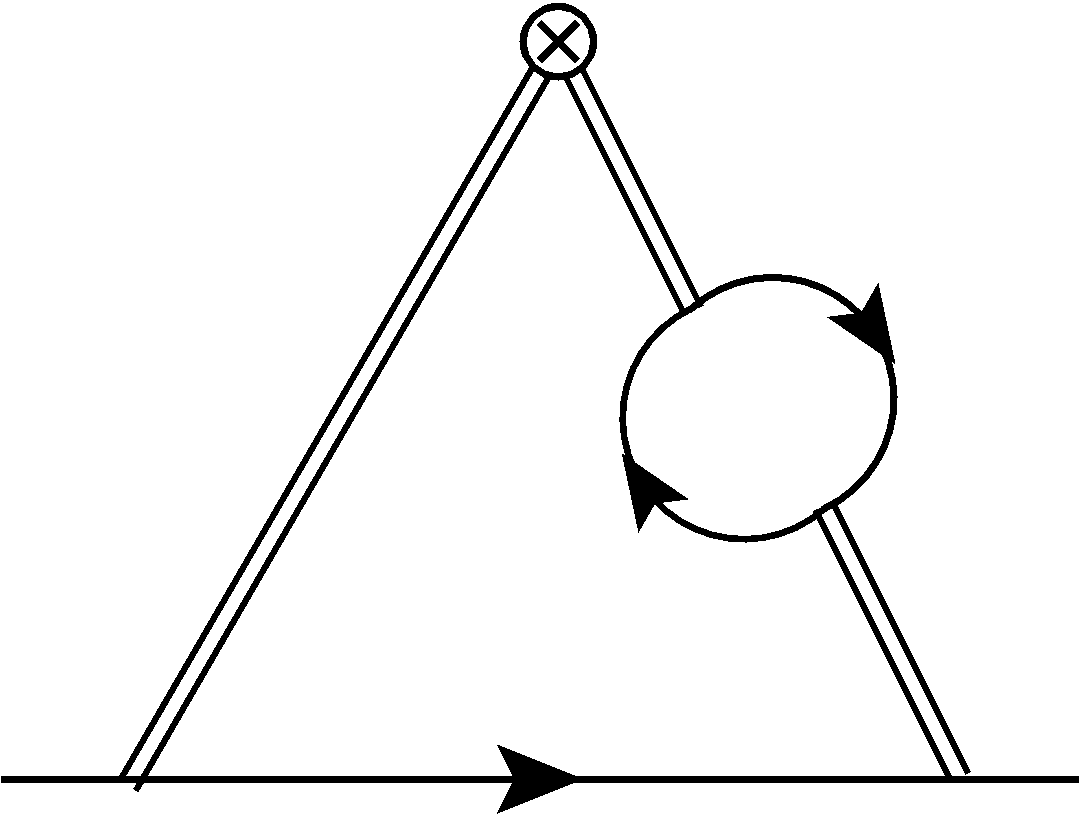}
		\label{img2}
	\end{minipage}
	\caption{\footnotesize{Diagrams that contribute to $\vev{O_2(x)W}$ and include hypermultiplet fields. In this figure gauge fields are denoted with wiggly line, scalar fields from the vector multiplet are denoted with double line, and hypermultiplet fields are denoted by a plain line (with an arrow for fermions and without an arrow for scalars).} }
\end{figure}

\subsection{Cusp anomalous dimension to three-loop order}

At order $g^6$, hypermultiplet fields appear in diagrams of two types; two-loop correction to the scalar and gauge field propagator (figure 2c) and one-loop correction to the vertex of three bosonic fields from the vector multiplet (figure 2d).  We will restrict to the case of $SU(2)$ gauge group  and we will compare the theory with four fundamental hypermultiplets ($\CN=2$ SQCD) to the one with one adjoint hypermultipet ($\CN=4$).  The one-loop correction for the vertex of three vector multiplet bosonic fields is the same for the two theories.\cite{Andree:2010na}  
 The diagramatic differences between the two-loop correction to the propagators in the two theories were calculated in \cite{Andree:2010na}.  It was shown that the two-loop propagator $D^{(2)}(x,y)$ of the gauge field or vector multiplet scalar satisfies
\begin{equation}
D^{(2)}(x,y)_{\CN=4}-D^{(2)}(x,y)_{\CN=2}={15\over {64\pi^4}}\zeta(3)g^4D^{(0)}(x,y)~.
\end{equation} 
This leads to 
\begin{equation}
\vev{W_\varphi}_{\CN=4}-\vev{W_\varphi}_{\CN=2}={15\over {64\pi^4}}\zeta(3)g^4\vev{W_\varphi}_{\CN=4}+\CO(g^8)~.\end{equation} 
Thus,
\begin{equation}
\begin{aligned}
B_{\CN=4}-B_{\CN=2}&={15\over {64\pi^4}}\zeta(3)g^4B_{\CN=4} +\CO(g^8)\\&={45\over {2048\pi^6}}\zeta(3)g^6 +\CO(g^8)\;,
\end{aligned}
\end{equation} 
where we have used $B_{\CN=4}={{3}\over{32\pi^2}}g^2+\CO(g^4)$ for a probe in the fundamental representation. 

To compare this with our conjecture, we use the localization result for the expectation value of a Wilson loop (on the ellipsoid) in the fundamental representation:\footnote{To get (\ref{Wb4-Wb2}) one has to multiply the ${\mathbb S}^4$ result (for example equation (11) in \cite{Andree:2010na}) by a factor of $b^2$ coming from the Wilson loop insertion.}
\begin{equation}
\left\langle W_b \right\rangle_{\CN=4}-\left\langle W_b \right\rangle_{\CN=2}={45\over {1024\pi^4}}\zeta(3)g^6b^2+\CO(g^8)\;.\label{Wb4-Wb2}
\end{equation}

Thus, according to our conjecture
\begin{equation}
\begin{aligned}
&B_{\CN=4}-B_{\CN=2}=\\&\frac{1}{4\pi ^2} \partial_b \big(\vev{W_b}_{\CN=4}-\vev{W_b}_{\CN=2} \big)|_{b=1}+\CO(g^8)=\\&
{45\over {2048\pi^6}}\zeta(3)g^6+\CO(g^8)\;,
\end{aligned}
\end{equation}
where we have used the fact that $\vev{W_{b=1}}=1+\CO(g^6)$. This agreement is encouraging. 
Note that we have not computed $h_W$ to three loops, but the agreement above between $B$ and our conjecture~\eqref{theBguess} also gives indirect evidence that~\eqref{B2h_Wagain} holds.

\begin{figure}
	\begin{minipage}{.15\linewidth}
		\includegraphics[width=\linewidth]{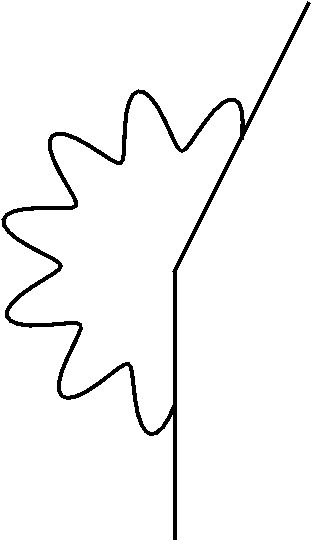}
		\subcaption{}
		\label{img1}
	\end{minipage}
	\hspace{.05\linewidth}
	\begin{minipage}{.15\linewidth}
		\includegraphics[width=\linewidth]{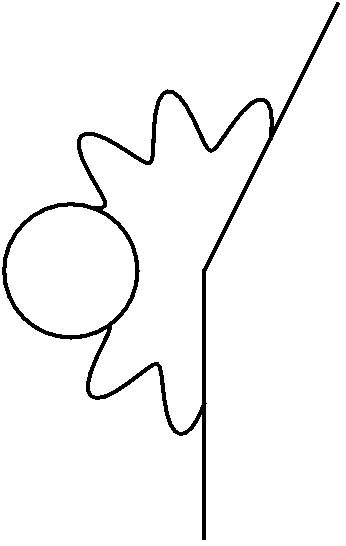}
		\subcaption{}
		\label{img2}
	\end{minipage}
	\hspace{.05\linewidth}
	\begin{minipage}{.15\linewidth}
		\includegraphics[width=\linewidth]{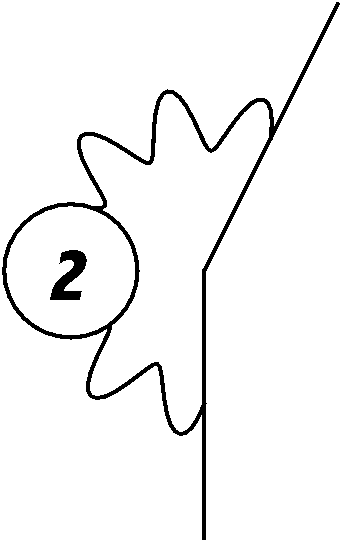}
		\subcaption{}
		\label{img1}
	\end{minipage}
	\hspace{.05\linewidth}
	\begin{minipage}{.15\linewidth}
		\includegraphics[width=\linewidth]{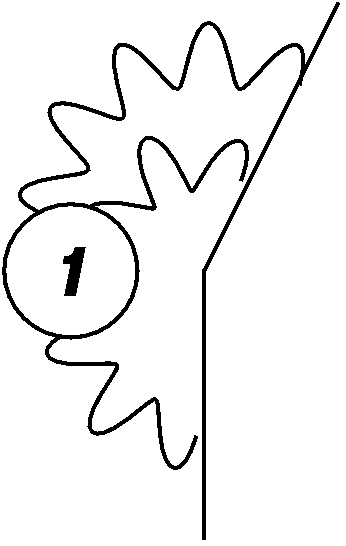}
		\subcaption{}
		\label{img2}
	\end{minipage}
	\caption{\footnotesize{Schematic plot of some of the Feynman diagrams that contribute to $\vev{W_{\varphi}}$. In this figure wiggly line is used to denote vector multiplet fields (scalars or vectors) and  a plain line is used to denote hypermultiplet fields (scalars or fermions):  (a) 1-loop diagrams. (b) 2-loop diagrams that involve hypermultiplet fields. (c)+(d) The 3-loop diagrams that involve hypermultiplet fields include the 2-loop correction to the propagator of the vector multiplet bosonic fields and the 1-loop correction to the vertex of three vector multiplet bosonic  fields.}}
	\end{figure}

\subsection{Positivity}
Since $B$ is by definition positive, the consistency of our proposal requires that the right hand side of (\ref{theBguess}) be positive. 

Let us check this claim for the case of $SU(N)$ with $N_f=2N$.
The derivative of the Wilson loop insertion,
\begin{equation}
f(b; a)\equiv\partial_b \hbox{Tr} e^{-2\pi b a}=\hbox{Tr}\left( (-2\pi a) e^{-2\pi a b}\right)\;,
\end{equation} 
is positive at $b=1$ since 
\begin{equation}
\partial_b f(b; a)= \hbox{Tr}\left((-2\pi a)^2 e^{-2\pi a b}\right)>0 \end{equation} due to the hermiticity of $a$. Therefore 
\begin{equation}
f(1; a)>f(0;  a)=-2\pi\hbox{Tr}(a)=0\;. 
\end{equation}

Since the classical, 1-loop, and instanton contributions are also positive, we obtain that 
\begin{equation}
\partial_b \,\ln \vev{W_b} |_{b=1}>0\;.
\end{equation}

\section{Additional Implications and Open Questions}
We end this article by pointing out two additional implications of the formula we have conjectured \eqref{theBguess}, and we suggest some open questions. 

The first implication concerns with the change in entanglement entropy due to the presence of a heavy probe. For any 4d CFT in its vacuum state the additional entanglement entropy of a spherical region due to the presence of a heavy probe located at its center is given by  \cite{Lewkowycz:2013laa}:
\begin{equation}
\begin{aligned}
S&=\log \vev{W}-8\pi^2h_W\;.
\end{aligned}\label{entanglement}
\end{equation}
Our conjecture~\eqref{theBguess} then implies that the additional entanglement due to a heavy quark in $\CN=2$ SCFTs is\footnote{ Written in this way, the formula is reminiscent of the formula in 3d CFTs of~\cite{Lewkowycz:2013laa},
\begin{equation}
S=\left(1-\frac{1}{2}\partial_b\right)\log \vev{W_b} |_{b=1}\hspace{0.7cm} 3d\;\; {\cal N}=2\;\; \hbox{SCFT}
\end{equation} 
A tempting guess is that for some class of $d$-dimensional CFTs $S=\left(1-\frac{d-2}{d-1}\partial_b\right)\log \vev{W_b} |_{b=1}$.
}
\begin{equation}
S=\left(1-\frac{2}{3}\partial_b\right)\log \vev{W_b} |_{b=1}~.\hspace{1cm} 
\end{equation}

The second implication concerns with the transcendentality of each term in the perturbative expansion of $B$ for $\CN=2$ SCFTs. Recall that in $\CN=4$ SYM, $\Gamma_{\text{cusp}}^\infty$ satisfies the rule of maximal transcendentality \cite{Kotikov:2002ab}: when expanded in powers of $g/\pi$, the coefficient of $(g/\pi)^{2n}$ has transcendentality $2n-2$. 
Interestingly, it follows immediately from (\ref{BinN=4}) that $B_{{\cal N}=4}$ also satisfies the same rule. \cite{Correa:2012at}

We now want to show that for $\CN=2$ SCFTs, the conjecture~(\ref{theBguess}) implies that $B_{\CN=2}$ obeys the following rule:  to each order in perturbation theory the leading transcendentality terms in the Bremsstrahlung function are given by the $\CN=4$ result. 

As far as the (perturbative) computation of $B$ goes, according to our conjecture, the difference between ${\mathcal{N}=4}$ and $\mathcal{N}=2$ lies in the non-trivial 1-loop determinant, which is given by 
\begin{equation}
Z_{\text{1-loop}}=\prod_{\alpha}H(ig\alpha\cdot \hat a)\prod_{\CR}\prod_{w_\CR}H(igw_\CR\cdot \hat a)^{-n_\CR}\;,
\end{equation}
where we changed the integration variable to ${\hat a=\frac{a}{g}}$. The function $H(x)$ is given in terms of the Barnes $G$-function as  $H(x)=G(1+x)G(1-x)$. The product runs over the roots ($\alpha$), the different representations ($\CR$), and the weights in each representation ($w_\CR$). The number of hypermultiplets in the representation $\CR$ is denoted by $n_\CR$. (For $\CN=4$, $Z_{\text{1-loop}}=1$.)

Using the expansion 
\begin{equation}
\log H(x)=-(1+\gamma)x^2-\sum_{n=2}^{\infty}\zeta(2n-1){{x^{2n}}\over n}\;,
\end{equation}
we find that
\begin{equation}
\begin{aligned}
&\log Z_{\text{1-loop}}=\\
&\sum_{n=2}^{\infty}{{\zeta(2n-1)}\over n}\left(\sum_\CR n_\CR\sum_{w_\CR}(iw_\CR\cdot \hat a)^{2n}-\sum_\alpha (i\alpha\cdot \hat a)^{2n}\right)g^{2n},
\end{aligned}
\end{equation}
where the $g^2$ term is absent due to conformal invariance. Thus, the $g^{2n}$ term in the expansion of $\log Z_{\text{1-loop}}$ (if non-vanishing) has degree of transcendentality $2n-1$. On the other hand, the $g^{2n}$ coefficient in the expansion of the Wilson loop insertion, $\hbox{Tr}\, e^{-2\pi \hat a g b}$ , has degree of transcendentality $2n$. Using this one can easily check that the leading transcendentality terms in (\ref{theBguess}) are not affected by the non-trivial 1-loop determinant. This proves our claim.
  
Finally, let us mention three open questions. One obvious question is to strengthen the body of perturbative evidence for the validity of~\eqref{theBguess}. For example, it would be very nice to consider order $g^6$ computations also for general gauge groups and not just $SU(2)$. A second question is to find a nontrivial check for the nonperturbative corrections to the cusp anomalous dimension entailed by~\eqref{theBguess}. (Indeed, unlike in $\mathcal{N}=4$ theories, if our conjecture is correct, then the cusp anomalous dimension in $\mathcal{N}=2$ theories also receives non-perturbative corrections.)  An additional question is to understand better the relation between derivatives with respect to the equivariant parameters and insertions of the energy-momentum supermultiplet. The relation may be nontrivial, for instance, due to the anomaly discussed in~\cite{Gomis:2015yaa}.

\appendix

\section{}\label{appendix}
In this appendix we show that a single function, $h_W$, determines the expectation values of all the bosonic operators in the stress-energy multiplet in the presence of the Wilson line.  

The reference we use for the supersymmetry transformations of the supercurrent multiplet is \cite{Fisher:1982fu}. The Weyl spinor indices conventions are those of Wess and Bagger. The $SU(2)_R$ indices are denoted by $i,j=1,2$ and are raised and lowered using an antisymmetric tensor which is denoted by $g$; $\psi^i=g^{ij}\psi_j$, $\psi_i=g_{ij}\psi^j$.

Consider the following Wilson line:
\begin{equation}
W_{\CR}={1\over {\text{dim} \CR}}\text{Tr}_\CR \CP\,\exp\; i \int(A_4+ \Phi+\bar{\Phi})dx^4\label{WilsonLine}\;.
\end{equation}
The supersymmetry transformations of the vector multiplet fields are \cite{Fisher:1982fu}:
\begin{equation}
\begin{aligned}
&\delta A_\mu=i\left(\bar\xi_{\dot{\alpha}i}\bar{\sigma}_\mu^{\dot{\alpha}\alpha}\lambda^{i}_\alpha-\bar{\lambda}_{\dot{\alpha}i}\bar{\sigma}^{\dot{\alpha}\alpha}_{\mu}\xi^i_{\alpha}\right)\;,\\
&\delta \Phi=g_{12}\lambda_i^{\alpha}\xi^i_\alpha\;,\\
&\delta \lambda_\alpha^i=F_{\mu\nu}{{\sigma^{\mu\nu}}_\alpha}^\beta\xi^i_\beta-2ig_{12}\sigma^{\mu}_{\alpha\dot \alpha}D_\mu\Phi\bar{\xi}^{\dot{\alpha}i}- 2ig[\bar\Phi,\Phi]\xi^i_\alpha\;.
\end{aligned}
\end{equation}
The combination $A_4+\Phi+\bar{\Phi}$ (and therefore the Wilson line (\ref{WilsonLine})) is invariant under supersymmetry transformations that satisfy \begin{equation}\bar{\xi}_i^{\dot{\alpha}}=-ig_{12}\bar{\sigma_4}^{\dot{\alpha}\alpha}\xi_{i\alpha}\;.\label{condition}
\end{equation}

In addition to $T_{\mu\nu}$ the supercurrent multiplet contains the supercurrents $J^{\mu i}_{\alpha}$, $\bar{J}^\mu_{i\dot{\alpha}}$, 	the $SU(2)_R$ current ${{t^\mu}_i}^j$, the $U(1)_R$-current $j^\mu$, a scalar operator $O_2$, fermionic fields $\chi^i_{\alpha}$, $\bar{\chi}^i_{\dot{\alpha}}$,  and spin (1,0) operator ${H_{\alpha}}^\beta$.

The supersymmetry transformations are given by \cite{Fisher:1982fu}:
\begin{equation}
\begin{aligned}
&{\delta O_2}=\bar\chi_{\dot{\alpha}i}\bar\xi^{\dot{\alpha}i}+\xi^\alpha_i\chi_\alpha^i\;,\\
&\delta\chi_\alpha^i={H_\alpha}^\beta\xi^i_\beta+{1\over2}\sigma^\mu_{\alpha\dot{\alpha}}j_\mu\bar{\zeta}^{\dot{\alpha}i}+{1\over2}{t_{\mu j}}^i{\sigma^\mu}_{\alpha\dot{\alpha}}\bar{\xi}^{\dot{\alpha}j}
\\&\;\;\;\;\;\;\;\;\;\;+i\sigma^{\mu}_{\alpha\dot{\alpha}}\partial_\mu O_2\bar\zeta^{\dot{\alpha}i}\;,\\
&\delta{H_\alpha}^\beta={i\over8}\left(J_{\mu i}^\beta\sigma^\mu_{\alpha\dot{\alpha}}\bar{\zeta}^{\dot{\alpha}i}-\bar{\zeta}_{\dot\alpha i}{\bar\sigma_\mu}^{\;\dot{\alpha}\beta}J_{\beta}^{\mu i}\right)\\
&\;\;\;\;\;\;\;\;\;\;\;\;\;+{{2i}\over3}\left(\bar{\xi}_{\dot{\alpha}i}{\bar{\sigma}}^{\mu\dot{\alpha}\beta}\partial_{\mu}\chi_\alpha^i-\partial_\mu\chi_i^\beta\sigma^\mu_{\alpha\dot{\alpha}}\bar{\xi}^{\dot{\alpha}}\right)\;,\\
&\delta j_\mu={i\over2}\left(J_{\mu i}^\alpha\xi^i_\alpha+\bar{J}_{\mu\dot{\alpha}i}\bar\xi^{\dot{\alpha}i}\right)\\&\;\;\;\;\;\;\;\;\;\;-{{8i}\over3}\left(\xi_i^\alpha{\sigma_{\mu\nu\alpha}}^\beta\partial^\nu\chi_\beta^i+\bar{\xi}_{\dot{\alpha}i}\bar{\sigma}_{\mu\nu\;\dot{\beta}}^{\;\;\;\dot{\alpha}}\partial^\nu\bar{\chi}^{\dot{\beta}i}\right)\;,\\
&\delta {t_{\mu i}}^j=i\left(J_{\mu i}^\alpha\xi_\alpha^j-\bar{\xi}_{\dot{\alpha}i}{\bar J}_\mu^{\dot{\alpha j}}\right)-{i\over2}\delta_i^j\left(J_{\mu k}^\alpha\xi_\alpha^k -\bar\xi_{\dot{\alpha}k}\bar J^{\dot{\alpha }k}_\mu\right)\\
&\;\;\;\;\;\;\;\;\;\;-{{4i}\over 3}\Big(\xi^\alpha_i{\sigma_{\mu\nu\alpha}}^\beta\partial^\nu\chi_\beta^j-\partial^\nu\chi^\alpha_i{\sigma_{\mu\nu\alpha}}^\beta\xi^j_\beta\\&\;\;\;\;\;\;\;\;\;\;+\partial^\nu{\bar{\chi}}_{\dot{\alpha}i}\bar{\sigma}_{\mu\nu\;\dot{\beta}}^{\;\;\;\dot{\alpha}}\bar{\zeta}^{\dot \beta j}-\bar{\zeta}_{\dot{\alpha}i}\bar{\sigma}_{\mu\nu\;\dot{\beta}}^{\;\;\;\dot{\alpha}}\partial^\nu\bar{\chi}^{\dot{\beta}j}\Big)\;,\\
&\delta J^{\mu i}_\alpha=2i\sigma_{\nu\alpha\dot{\alpha}}\bar{\xi}^{\dot \alpha i}T^{\mu\nu}\\
&\;\;\;\;\;\;\;\;\;\;\;-4\left(\partial_\nu {H_\alpha}^\beta{{\sigma^{\mu\nu}}_\beta}^\gamma+{1\over3}{{\sigma^{\mu\nu}}_\alpha}^\beta\partial_\nu{H_\beta}^\gamma\right)\xi^i_\gamma\\
&\;\;\;\;\;\;\;\;\;\;+{1\over3}\left({\sigma^{\mu\nu}}{\sigma^{\lambda}}-3\sigma^{\lambda}{\bar{\sigma}}^{\mu\nu}\right)_{\alpha\dot{\alpha}}\bar{\xi}^{\dot{\alpha}j}\left(\delta^i_j\partial_\nu j_\lambda-2\partial_\nu {t_{\lambda j}}^i \right)\;,\\
&\delta T_{\mu\nu}={1\over2}\xi^\alpha_i{{\sigma^{\mu\lambda}}_\alpha}^\beta\partial_\lambda J_\beta^{\nu i}-{1\over2}\bar{\xi}_{\dot \alpha i}{\bar{\sigma}^{\mu\lambda\dot{\alpha}}}_{\;\;\;\;\;\;\dot{\beta}}\partial_\lambda\bar J^{\nu\dot \beta i}+\mu\leftrightarrow\nu\;.
\end{aligned}
\end{equation}

For transformation parameters that satisfy (\ref{condition}), we have the Ward identity
\begin{equation}
\left\langle W\delta G \right\rangle=\left\langle \delta (W G) \right\rangle=0\;,
\end{equation}
for any operator $G$.
The Wilson line is an $SU(2)_R$ singlet and therefore ${\left\langle W {{t_{\mu i}}}^j(x)\right\rangle=0}$. In addition, the symmetries of the Wilson line configuration together with the conservation of the $U(1)_R$ current imply that ${\left\langle W j_{\mu}(x)\right\rangle=0}$. Using these, the Ward identity ${\left\langle W\delta \chi^i_\alpha \right\rangle
=0}$ becomes
\begin{equation}
\left\langle {H_\alpha}^\beta \right\rangle_W=-g_{12}{(\sigma^\mu\bar{\sigma}_4)_\alpha}^\beta\partial_\mu \left\langle O_2 \right\rangle_W \;.
\end{equation}
Plugging all the above into $\left\langle W \delta J^{\mu i}_\alpha \right\rangle
=0$ we get
\begin{equation}\begin{aligned}
{(\sigma_\nu\bar{\sigma}_4)}\left\langle T^{\mu\nu} \right\rangle_W= -2\left(\sigma^\rho\bar\sigma^4\sigma^{\mu\nu}+{1\over3}\sigma^{\mu\nu}\sigma^\rho\bar\sigma^4\right)\partial_\nu\partial_\rho\left\langle O_2  \right\rangle_W,\\
\end{aligned}
\end{equation}
or:
\begin{equation}
\left\langle T^{\lambda\mu} \right\rangle_W= \text{Tr}\left({1\over3}\sigma^\rho\bar{\sigma}^\lambda\sigma^{\mu\nu}-\sigma^4\bar{\sigma}^\lambda\sigma^\rho\bar\sigma^4\sigma^{\mu\nu}
\right)\partial_\nu\partial_\rho\left\langle O_2  \right\rangle_W\;.
\end{equation}

Using that $\partial_4\left\langle O_2(x)\right\rangle_W=0$, and the properties of the $\sigma$-matrices this becomes
\begin{equation}
\begin{aligned}
&\left\langle T_{ab}(x) \right\rangle_W=-{2\over3}\left(\delta_{ab}\partial^2-\partial_a\partial_b\right)\left\langle O_2(x) \right\rangle_W\;,\\
&\left\langle T_{a4}(x) \right\rangle_W=0\;,\\
&\left\langle T_{44}(x) \right\rangle_W={4\over3}\partial^2\left\langle O_2(x) \right\rangle_W\;,
\end{aligned}
\end{equation}
where $a,b=1,2,3$. 
Given that $\left\langle O_2(x) \right\rangle_W={C\over l^2}$, with $l^2=x^ax^a$ and $C$ a constant, we finally obtain 
\begin{equation}
\begin{aligned}
&\left\langle T_{ab}(x) \right\rangle_W=-{{h_{W}}\over l^4}\left(\delta_{ab}-{{2x_ax_b}\over l^2}\right)\;,\\
&\left\langle T_{a4}(x) \right\rangle_W=0\;,\\
&\left\langle T_{44}(x) \right\rangle_W={{h_{W}}\over l^4}\;,
\end{aligned}
\end{equation}
with $$h_W={8\over3}C\;.$$

\begin{acknowledgements}
We would like to thank O. Aharony, J. Gomis, J.~Henn,  A.~Lewkowycz, J. Maldacena, V. Mitev, and E. Pomoni for discussions. B.F. would like to thank the Weizmann Institute of Science for hospitality. E.G. and Z.K. are supported by the ERC STG grant 335182, by the Israel Science Foundation under grant 884/11, by the United
States-Israel Bi-national Science Foundation (BSF) under grant 2010/629, by the
Israel Science Foundation center for excellence grant (grant no.1989/14) and also by the  I-CORE Program of the Planning and Budgeting Committee. B.F. is partially funded by the Spanish MINECO under projects FPA2013-46570-C2-2-P and MDM-2014-0369 of ICCUB (Unidad de Excelencia 'María de Maeztu') and by AGAUR, grant 2014-SGR-1474. 

\end{acknowledgements}

\end{document}